\documentclass[aps,prb,twocolumn,superscriptaddress,showpacs]{revtex4}

\usepackage{graphicx}

\begin{document}


\title{Short-range magnetic correlations in Tb$_{5}$Ge$_{4}$}



\author{W. Tian}
\affiliation{Department of Physics and Astronomy, Iowa State
University, Ames, Iowa 50011, USA} \affiliation{Ames Laboratory, US
DOE, Iowa State University, Ames, Iowa 50011, USA}

\author{A. Kreyssig}
\affiliation{Department of Physics and Astronomy, Iowa State
University, Ames, Iowa 50011, USA} \affiliation{Ames Laboratory, US
DOE, Iowa State University, Ames, Iowa 50011, USA}

\author{J. L. Zarestky}
\affiliation{Department of Physics and Astronomy, Iowa State
University, Ames, Iowa 50011, USA} \affiliation{Ames Laboratory, US
DOE, Iowa State University, Ames, Iowa 50011, USA}

\author{L. Tan}
\affiliation{Department of Physics and Astronomy, Iowa State
University, Ames, Iowa 50011, USA} \affiliation{Ames Laboratory, US
DOE, Iowa State University, Ames, Iowa 50011, USA}

\author{S. Nandi}
\affiliation{Department of Physics and Astronomy, Iowa State
University, Ames, Iowa 50011, USA} \affiliation{Ames Laboratory, US
DOE, Iowa State University, Ames, Iowa 50011, USA}

\author{A. I. Goldman}
\affiliation{Department of Physics and Astronomy, Iowa State
University, Ames, Iowa 50011, USA} \affiliation{Ames Laboratory, US
DOE, Iowa State University, Ames, Iowa 50011, USA}

\author{T. A. Lograsso}
\affiliation{Ames Laboratory, US DOE, Iowa State University, Ames,
Iowa 50011, USA}

\author{D. L. Schlagel}
\affiliation{Ames Laboratory, US DOE, Iowa State University, Ames,
Iowa 50011, USA}

\author{K. A. Gschneidner}
\affiliation{Ames Laboratory, US DOE, Iowa State University, Ames,
Iowa 50011, USA} \affiliation{Department of Materials Science and
Engineering, Iowa State University, Ames, Iowa 50011, USA}

\author{V. K. Pecharsky}
\affiliation{Ames Laboratory, US DOE, Iowa State University, Ames,
Iowa 50011, USA} \affiliation{Department of Materials Science and
Engineering, Iowa State University, Ames, Iowa 50011, USA}

\author{R. J. McQueeney}
\affiliation{Department of Physics and Astronomy, Iowa State
University, Ames, Iowa 50011, USA} \affiliation{Ames Laboratory, US
DOE, Iowa State University, Ames, Iowa 50011, USA}

\begin{abstract}

We present a single crystal neutron diffraction study of the
magnetic short-range correlations in Tb$_5$Ge$_4$ which orders
antiferromagnetically below the Neel temperature $T_N$ $\approx$ 92
K. Strong diffuse scattering arising from magnetic short-range
correlations was observed in wide temperature ranges both below and
above $T_N$. The antiferromagnetic ordering in Tb$_5$Ge$_4$ can be
described as strongly coupled ferromagnetic block layers in the
$ac$-plane that stack along the $b$-axis with weak antiferromagnetic
inter-layer coupling. Diffuse scattering was observed along both
$a^*$ and $b^*$ directions indicating three-dimensional short-range
correlations. Moreover, the $q$-dependence of the diffuse scattering
is Squared-Lorentzian in form suggesting a strongly clustered
magnetic state that may be related to the proposed Griffiths-like
phase in Gd$_5$Ge$_4$.
\end{abstract}

\pacs{valid numbers to be inserted here
      }

\maketitle

\section{Introduction}

Tb$_5$Ge$_4$ and Gd$_5$Ge$_4$ belong to the rare earth
R$_5$(Si$_x$Ge$_{1-x}$)$_4$ series compounds. These materials
exhibit large magnetocaloric (MC) effects \cite{1Pecharsky1997,
2Pecharsky1997, 3Pecharsky1997, 4Pecharsky1997} and are currently
attracting much attention for their potential application in
magnetic refrigeration
\cite{2Pecharsky1997,5Pecharsky1998,6Miller2006,7Morellon2001}. Both
Tb$_5$Ge$_4$ and Gd$_5$Ge$_4$ are rich in magnetic properties
\cite{CALB2005,TPGP2004,LGP2002,OPG2006,
OuyangPRB2006,LevinPRB2004,MAM03,RitterPRB2002,MagenPRL2006} and are
believed to play a key role in understanding the underlying physics
of the R$_5$(Si$_x$Ge$_{1-x}$)$_4$ systems. They both crystallize in
the Sm$_5$Ge$_4$-type crystallographic structure and adopt the same
magnetic space group, $Pnm'a$ \cite{SP78,RitterPRB2002,TANPRB2005}.
Tb$_5$Ge$_4$ and Gd$_5$Ge$_4$ undergo long range antiferromagnetic
(AFM) transitions at $\sim$ 92 K and $\sim$ 127 K, respectively. The
magnetic structure of Tb$_5$Ge$_4$ and Gd$_5$Ge$_4$ consist of
Tb/Gd-rich block layers in the $ac$-plane that stack along the
$b$-axis with strong ferromagnetic intralayer and weak AFM
interlayer interactions. Gd$_5$Ge$_4$ has a collinear AFM structure
with the magnetic moments lying within the block layers along the
c-axis. Tb$_5$Ge$_4$ orders in a similar fashion, although the
single-ion anisotropy results in significant canting of the moments
at low temperature.

\begin{figure}[!t]
\centering\includegraphics[width=3.2in]{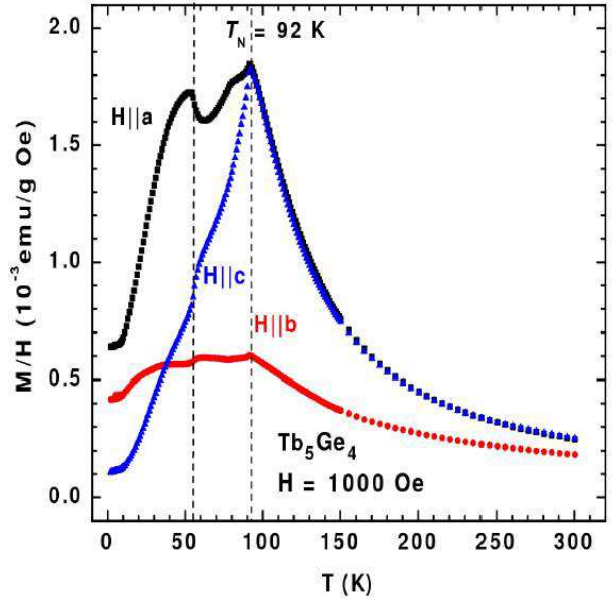}
\caption{\label{fig:chi}(Color online) Temperature dependencies of
the magnetic susceptibility measured with applied magnetic field (H
= 1000 Oe, zero field cooled) parallel to all three crystallographic
axes.}
\end{figure}

\begin{figure*}
\centering\includegraphics[width=6.5in]{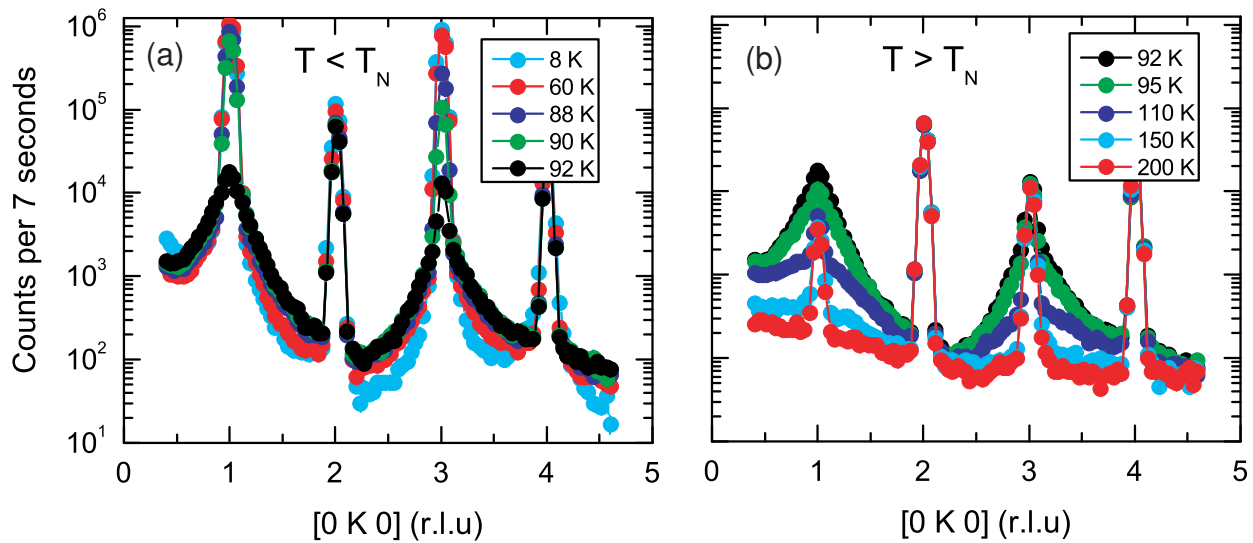}
\caption{\label{fig:longscan}(Color online) Longitudinal scans (in
Log10 scale) along the (0$k$0) direction measured at several
temperatures (a) below $T_N$, (b) above $T_N$, indicate strong AFM
SRC with broad magnetic scattering observed around $k$ = odd. The
remaining scattering intensity of (010) and (030) at 200 K is likely
due to multiple scattering.}
\end{figure*}

Figure \ref{fig:chi} shows the magnetic susceptibility of
Tb$_5$Ge$_4$ along all three crystallographic axes measured with a
Quantum Design SQUID Magnetic Properties Measurement System. Two
phase transitions were observed: the AFM transition at $T_N \sim$ 92
K and a second phase transition at $\sim$ 55 K. The $\sim$ 55 K
transition was attributed to a spin reorientation transition similar
to that proposed for Gd$_5$Ge$_4$ at $\sim$ 75 K \cite{TANthesis}.
It has been suggested that the spin reorientation transition in both
Gd$_5$Ge$_4$ and Tb$_5$Ge$_4$ may arise from the delicate
competition between the magnetic anisotropy from the spin-orbit
coupling of the conduction electrons and the dipolar interaction
anisotropy \cite{TANthesis}. In addition to the AFM and spin
reorientation transitions, it is of particular interest that
significant magnetic short-range correlations (SRC) was suggested in
Gd$_5$Ge$_4$ at temperatures both below and above $T_N$ based upon
the low magnetic field dc magnetization and ac magnetic
susceptibility measurements \cite{OuyangPRB2006}. It has been
interpreted as evidence of a Griffiths-like phase similar to the one
observed in Tb$_5$Si$_2$Ge$_2$, investigated by small-angle neutron
scattering \cite{MagenPRL2006}. A Griffiths phase (GP) \cite{Gri69}
is a nanoscale magnetic clustering phenomenon that is usually
associated with competing magnetic interactions in the system
\cite{GriPRL2005, GriPRL2002, GriPRL1998}. It is possible for a
Griffiths-like phase to exist in Gd$_5$Ge$_4$ due to the competition
between FM and AFM interactions in Gd$_5$Ge$_4$, FM interactions
within the layers, and either AFM or FM interactions between the
layers (small structural changes or applying relatively low magnetic
field comparing to $T_N$ can switch the interlayer interactions
between AFM \& FM interactions therefore switch the low temperature
phase between AFM order or FM order). However, studies of
Gd$_5$Ge$_4$ have been hampered by the large neutron absorption
cross-section of gadolinium. Tb$_5$Ge$_4$ exhibits similar magnetic
properties as to Gd$_5$Ge$_4$ hence it is an ideal candidate for
neutron scattering studies of magnetic SRC displayed in these
compounds. We report here neutron diffraction studies of the AFM
phase transition and the magnetic SRC in Tb$_5$Ge$_4$.

\section{Experimental Details}

A large Tb$_5$Ge$_4$ single crystal ($\sim$ 5 grams) was used for
the neutron diffraction experiment. The single crystal was grown at
the Materials Preparation Center \cite{material-center} using the
Bridgman technique as described in Ref.
\onlinecite{Bridgman-technique}. The mosaic of the crystal is
0.41(3)$^\circ$ along the $a$-axis and 0.63(2)$^\circ$ along the
$b$-axis as determined by the full width at half maximum of the
(400) and (060) Bragg peak rocking curves. The crystal was mounted
on a thin aluminum post, oriented in the ($hk$0) scattering plane,
and sealed in a helium filled aluminum sample can. A closed-cycle
Helium refrigerator (Displex) was used which allows accurate
temperature control between 10 K and 300 K. The experiments were
performed using the HB1A triple-axis spectrometer located at the
High Flux Isotope Reactor (HFIR) at the Oak Ridge National
Laboratory (ORNL). The HB1A spectrometer operates with a fixed
incident energy, \textit{E$_i$} = 14.6 meV using a double pyrolitic
graphite (PG) monochromator system. Two highly oriented PG filters
were mounted before and after the second monochromator to
significantly reduce higher order contaminations of the incident
beam (i.e., $I_{\lambda /2}\cong$ 10$^{-4}I_{\lambda })$. A
collimation of $open$-$40'$-$sample$-$40'$-$68'$ was used throughout
the experiment. All data have been normalized to the beam monitor
count.

\section{Results and Discussions}

Figure 2 compares longitudinal scans measured along the (0$k$0)
direction at selected temperatures both below
(Fig.~\ref{fig:longscan} (a)) and above (Fig.~\ref{fig:longscan}
(b)) $T_N$. Strong AFM magnetic reflections with $k$ = odd integer
were observed below $T_N$  consistent with the magnetic structure of
Tb$_5$Ge$_4$. At low temperatures, the (010) and (030) magnetic
reflections are superimposed on weak and very broad
Lorentzian-shaped diffuse scattering peaks arising from magnetic
fluctuations. Below $T_N$, the diffuse scattering increases with
increasing temperatures as illustrated in Fig.~\ref{fig:longscan}
(a). The diffuse scattering is strongest at $T_N$ and then weakens
as the temperature increases above $T_N$ as shown in
Fig.~\ref{fig:longscan} (b) as expected for typical critical
behavior. The data indicate strong diffuse scattering around the
strong magnetic reflections over a wide temperature range.

\begin{figure}[!t]
\centering\includegraphics[width=3.0in]{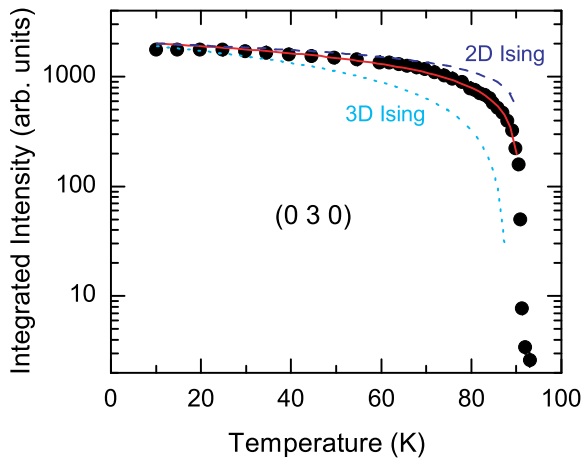}
\caption{\label{fig:order}(Color online) Tb$_5$Ge$_4$ order
parameter, integrated intensity of the (030) magnetic peak as a
function of temperature. The red line is the fitting of the order
parameter data to the power law as described in the text which
yields T$_N$ $\approx$ 91.38 $\pm$ 0.05 K and $\beta$ $\approx$ 0.20
$\pm$ 0.01, and the dashed and dotted lines are calculations using
the same fitting parameters but simply replacing the $\beta$ value
to be the values of a 2D Ising ($\beta$=0.125) and a 3D Ising
($\beta$=0.326) system.}
\end{figure}

We characterized the $T_N \sim$ 92 K AFM transition with an order
parameter measurement. Fig.~\ref{fig:order} depicts the order
parameter of Tb$_5$Ge$_4$ as measured by monitoring the strong
magnetic reflection (030) as a function of temperature. The
integrated intensity was obtained by fitting the (030) rocking curve
measured at each temperature to a Lorentzian function with a
constant background. A fit of the order parameter to a power-law
\textit{I(T)=I$_0$[(T$_N$-T)/T$_N$)]$^{2\beta}$} yields $T_N$
$\approx$ 91.38 $\pm$ 0.05 K and $\beta$ $\approx$ 0.20 $\pm$ 0.01,
where $\beta$ is the critical exponent. The fitting result is
plotted in Fig.~\ref{fig:order}, solid red line, in comparison to
calculations using the same fitting parameters but replacing $\beta$
with the values of a two-dimensional (2D) Ising ($\beta$=0.125) and
a three-dimensional (3D) Ising ($\beta$=0.326) system
\cite{Collins-book}. The obtained critical temperature $T_N$ is in
good agreement with the magnetic susceptibility result. The yielded
$\beta$ value $\sim$ 0.20 is between the theoretical values of a 2D
and a 3D Ising system, which suggests that the dimensionality of
Tb$_5$Ge$_4$ is intermediate between 2D and 3D consistent with its
layered structure.

\begin{figure} [!t]
\centering\includegraphics[width=2.8in]{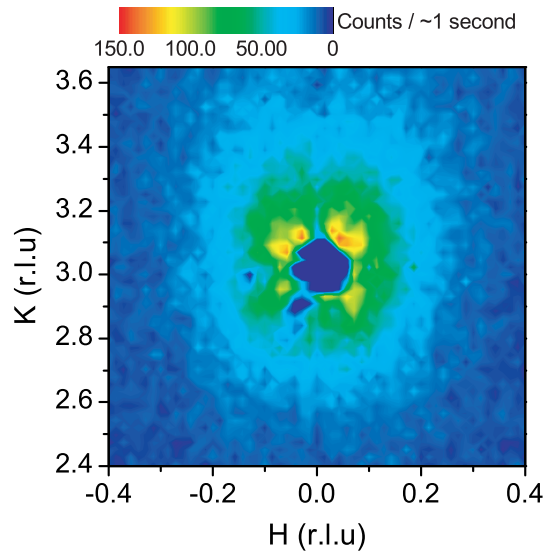}
\caption{\label{fig:mesh030}(Color online) Diffuse scattering
intensity (88 K - 8 K) vs. $k$ and $h$ around (030) constructed from
a series of grid scans measured at 8 K and 88 K. The 8 K data is
subtracted from the 88 K data to eliminate contributions from (030)
magnetic Bragg reflection.}
\end{figure}

The longitudinal scans shown in Fig.~\ref{fig:longscan} indicate
strong diffuse scattering along the $b$-axis. In order to see how
the diffuse scattering is distributed in the ($hk$0) plane, a series
of grid scans around (030) were performed at two temperatures, 8 K
and 88 K. The 8 K data were subtracted from the 88 K data to
eliminate contributions from the (030) magnetic Bragg reflection.
Fig.~\ref{fig:mesh030} is the contour plot of the subtracted diffuse
scattering intensity vs. $h$ and $k$. It shows that the diffuse
scattering extends along both $h$ and $k$ directions. The diffuse
scattering intensity is strong around ($\sim$0.1 $\sim$3.1 0) and
its equivalent positions. No strong anisotropy is observed
suggesting the magnetic correlations associated to the diffuse
scattering is not restricted to the FM block layer ($ac$-plane) but
is rather more three-dimensional exhibiting AFM critical scattering
behavior.

\begin{figure*}
\centering\includegraphics[width=5.8in]{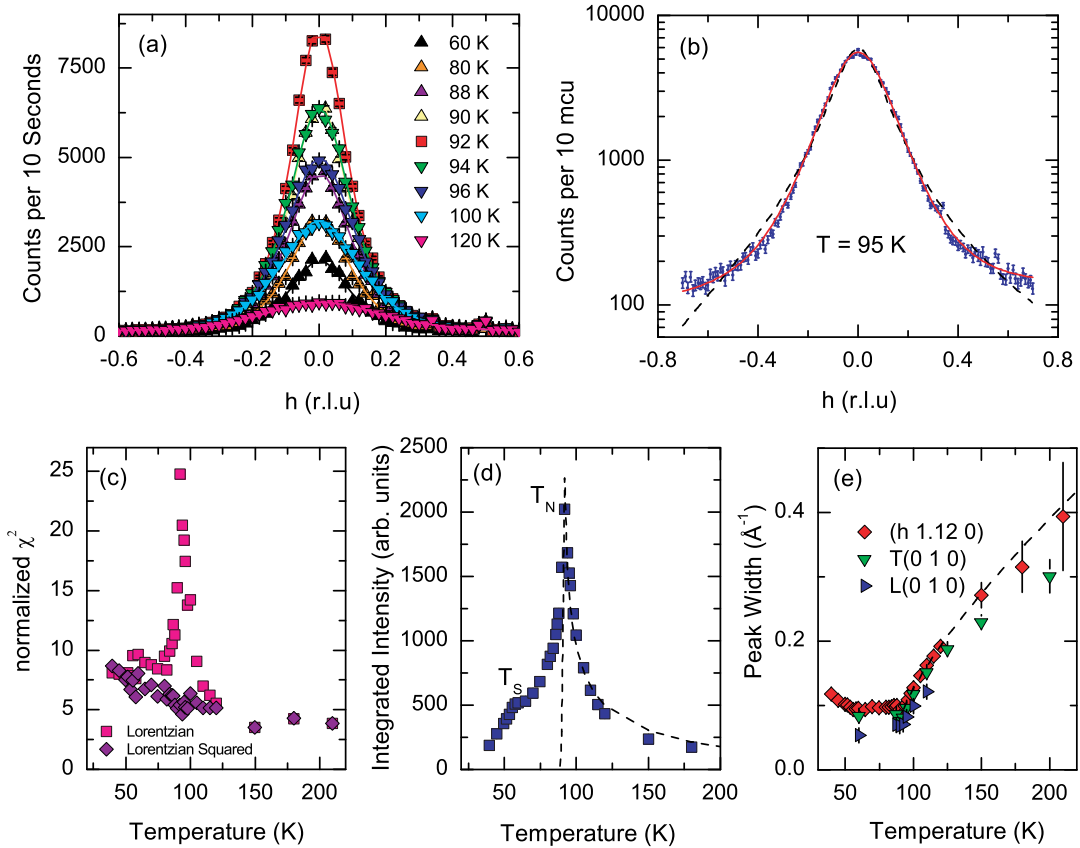}
\caption{\label{fig:diffuse}(Color online) Diffuse scattering along
the $h$-direction measured at (0 1.12 0). (a) Representative scans
measured at different temperatures. The solid curves are
least-squares fits to a Squared-Lorentzian function of $h$ as
described in the text; (b) Comparison of the fits of the 95 K data
to a Lorentzian + BG (blue dashed line) function only and a
Squared-Lorentzian + BG (red solid line) function only; (c) $\chi^2$
vs. T, where the normalized $\chi^2$ is obtained from least-squares
fits to a Lorentzian + BG only and a Squared-Lorentzian + BG
function only; (d) Integrated intensity vs. temperature; (e) Peak
width vs. temperature; The dashed black lines in (d) and (e) are
fits of the $T > T_N$ data to a power law as described in the text.}
\end{figure*}

The diffuse scattering was studied in detail as a function of
temperature. Wide transverse scans along the $h$ direction were
performed at (0 1.12 0) to reduce contributions from the nearby
(010) strong magnetic reflection. At 10 K, the FWHM (full width at
half maximum) peak width of the (010) is about 0.09 [r.l.u]
(reciprocal lattice unit). Therefore we may attribute the scattering
intensity measured at (0 1.12 0) to diffuse scattering from magnetic
correlations. Fig.~\ref{fig:diffuse} (a) shows typical scans at
different temperatures. The scattering intensity of the strong
diffuse scattering peak first increases with increasing temperature
up to $T_N$ and then decreases with further increasing temperature
above $T_N$. In general, the $q$-dependence (here $q$ = $h$ or $k$)
of diffuse scattering can be well described by the following
equation \cite{Motoya, Bentley}, a sum of a Lorentzian function and
a Squared-Lorentzian function plus constant background (BG) ,
\begin{equation}
{\mathcal I(q)} = \frac{A}{\kappa_L^2+q^2}
+\frac{B}{(\kappa_{LS}^2+q^2)^2} + BG. \label{eq:Lorentzian}
\end{equation}

The first Lorentzian term is the conventional critical scattering
component representing an Ornstein-Zernike form, i.e. exp$^{-\kappa
r}$/$r$, for the magnetic correlations \cite{Collins-book}. The
second Squared-Lorentzian term \cite{lovesey} is generally
considered to arise from static or frozen spin clusters within which
the spin correlations decrease more gradually as exp$^{-\kappa r}$.
Surprisingly, by fitting the data to a Lorentzian + BG function
only, a Squared-Lorentzian + BG function only, and the sum of
Lorentzian plus Squared-Lorentzian function as described in
Eq.~\ref{eq:Lorentzian}, respectively, we found that the diffuse
scattering data can be best described by the Squared-Lorentzian
function only plus constant background. Adding an extra Lorentzian
term does not improve the quality of the least-squares fit to data.
The comparisons between the fits to a Lorentzian only and a
Squared-Lorentzian only are shown in Fig.~\ref{fig:diffuse}. Note
that the diffuse scattering peak is very broad, thus the resolution
effect can be neglected and no resolution corrections are applied in
the data analysis. Fig.~\ref{fig:diffuse} (b) compares the fits of
the 95 K data to Lorentzian + BG (blue dashed line) only and
Squared-Lorentzian + BG (red solid line) only. This indicates that
the 95 K data can not be described by a Lorentzian + BG function. On
the other hand, as shown in Fig.~\ref{fig:diffuse} (a), the solid
curves are fits to a Squared-Lorentzian + BG only that adequately
describe the data for the measured $q$-range at all temperatures.
The comparison of the obtained normalized $\chi^2$ from these two
fittings is shown in Fig.~\ref{fig:diffuse} (c). It clearly shows
that the data are better captured by a Squared-Lorentzian form, the
Lorentzian line shape does not give a good description of the data,
particularly at temperatures near $T_N$ as illustrated in
Fig.~\ref{fig:diffuse} (b).

The integrated intensity and the FWHM obtained from least-squares
fits to the data with a Squared-Lorentzian function are plotted in
Fig.~\ref{fig:diffuse} (d) and (e). Two features are observed in the
integrated intensity data (Fig.~\ref{fig:diffuse} (d)). The small
kink at $\sim$ 55 K is associated with the spin reorientation
transition, and the peak at $\sim$ 92 K is associated with the AFM
transition. Both temperatures agree well with the magnetic
susceptibility data. Despite the small kink at $\sim$ 55 K, the
$\sim$ 92 K peak is nearly symmetric indicating strong critical
fluctuations at $T_N$ that die off as one moves away from $T_N$ in
either direction. Above $T_N$, the integrated intensity data can be
fit to a power law
\textit{I(T)=I$_0$[(T-T$_N$)/T$_N$)]$^{-2\beta^{\prime}}$} yielding
$T_N$ $\approx$ 89.94 $\pm$ 1 K and $\beta^{\prime}$ $\approx$ 0.22
$\pm$ 0.02 (dashed line in Fig.~\ref{fig:diffuse} (d)). The obtained
$\beta^{\prime}$ value agrees to the $\beta$ value obtained from the
fit to the order parameter data. As illustrated in
Fig.~\ref{fig:diffuse} (e), the correlation length of the SRC
remains relatively constant below $T_N$ as indicated by a nearly
constant peak width. Above $T_N$, the peak width increases as
expected as the correlation length decreases with increasing
temperature. A fit to the $T
> T_N$ peak width data to a power law
\textit{$\xi$(T)=$\xi$$_0$[(T-T$_N$)/T$_N$)]$^{\upsilon}$} gives
$T_N$ $\approx$ 91.7 $\pm$ 1 K and $\upsilon$ $\approx$ 0.77 $\pm$
0.06 (dashed line in in Fig.~\ref{fig:diffuse} (e)), the $\upsilon$
value is between the theoretical values of a 3D Ising ($\upsilon$ =
0.6312) and a 2D Ising ($\upsilon$ = 1) \cite{Collins-book}
consistent with the order parameter measurement results. The
Squared-Lorentzian peak widths obtained from fits to (010)
longitudinal and transverse scans are also shown in
Fig.~\ref{fig:diffuse} (e). The Lorentzian-squared lineshape
provides the best fit along both the $h$ and $k$ directions,
indicating that the correlations in the spin clusters extend both in
the block layers and between the blocks.

Our neutron diffraction study reveals strong diffuse scattering in
Tb$_5$Ge$_4$ that persists to temperatures well above $T_N$. A
detailed study of the peak shape indicates it is not conventional
critical scattering with a Lorentzian shape but shows a
Squared-Lorentzian peak shape. As described in Ref.
\onlinecite{lovesey} and Ref. \onlinecite{Cowley}, the
Squared-Lorentzian term arises if the pair correlation function
falls off as exp$^{-\kappa r}$, which is characteristic of a
spin-cluster state. Although the diffuse scattering is
Squared-Lorentzian in form providing evidence of a clustered
magnetic state in Tb$_5$Ge$_4$, we believe that the diffuse
scattering observed in Tb$_5$Ge$_4$ is quite different from the
proposed FM Griffiths-like phase in Gd$_5$Ge$_4$ (inferred from
dc/ac magnetization and magnetic suscpetibility studies)
\cite{OuyangPRB2006} for the following reasons. (1) As depicted in
Fig.~\ref{fig:longscan}, at temperatures both below and above $T_N$,
the diffuse scattering is peaked at odd values of $k$ (AFM
wavevector) only, indicating that it is associated with AFM
fluctuations. (2) The integrated intensity of the diffuse scattering
also behaves like that typical of magnetic critical fluctuations,
with a divergence of the correlation length at $T_N$. (3) The
critical exponents obtained by fits of the diffuse scattering
integrated intensity and peak width to a power law are consistent
with the values obtained from AFM order parameter measurements. (4)
Quasi-elastic measurements indicate the diffuse scattering is static
in origin. Our neutron diffraction data indicate that the diffuse
scattering observed in Tb$_5$Ge$_4$ exhibits behaviors of AFM
critical fluctuations despite the Squared-Lorentzian peak shape.

The fact that the peak shape of the diffuse scattering is not a
Lorentzian, as expected for normal critical scattering, is
interesting and should not be left without a discussion. Here we
consider two possibilities that may affect the diffuse scattering
peak shape. (1) The Squared-Lorentzian peak shape may be intrinsic,
i.e. related to the Griffiths-like phase, formation of which has
been discussed in Refs. \onlinecite{OuyangPRB2006} and
\onlinecite{MagenPRL2006}; (2) The unusual peak shape may also arise
from some extrinsic effects, for example impurities in Tb$_5$Ge$_4$.
It has been reported that R$_5$(Si$_x$Ge$_{1-x}$)$_3$-type impurity
phases, seen as very thin plates that are scattered through the bulk
of R$_5$(Si$_x$Ge$_{1-x}$)$_4$ samples, are present in all studied
compounds of this series regardless of R \cite{impurity}. Our data
show that Tb$_5$Ge$_3$ impurity phase is also present in the studied
Tb$_5$Ge$_4$ crystal. It is possible that the Tb magnetic
sublattices of Tb$_5$Ge$_4$ are disrupted by the Tb$_5$Ge$_3$
impurities resulting in spin-clusters in Tb$_5$Ge$_4$ which give
rise to the Squared-Lorentzian diffuse scattering peak shape.
\\
\\
\textbf{Acknowledgments}
\\
\\
Ames Laboratory is operated for the U.S. Department of Energy by
Iowa State University under Contract No. DE-AC02-07CH11358. The HFIR
is a national user facility funded by the United States Department
of Energy, Office of Basic Energy Sciences, Materials Science, under
Contract No. DE-AC05-00OR22725 with UT-Battelle, LLC.

\end{document}